\documentstyle[aaspp4]{article}

\received{26 December 1995}

\slugcomment{To appear in The Astrophysical Journal Letters, 1996}

\begin{document}

\def\fd{IRAS F10214}
\def\fz{IRAS 09104}
\def\fq{IRAS F15307}
\def\mmu{$m^{-1}$}
\def\mmum{$m^{-1/2}$}
\def\hmd{$h^{-2}$}
\def\hmu{$h^{-1}$}

\title{ DUST ENSHROUDED AGN MODELS FOR HYPERLUMINOUS HIGH
REDSHIFT IR GALAXIES}

\author{Gian Luigi\ Granato\altaffilmark{1}}
\affil{Osservatorio Astronomico di Padova, Padova, Italy}

\author{Luigi\ Danese\altaffilmark{2}}
\affil{International School for Advanced Studies}

\and

\author{Alberto Franceschini\altaffilmark{3}}
\affil{Dipartimento di Astronomia di Padova, Padova, Italy}

\altaffiltext{1}{Osservatorio Astronomico, Vicolo dell'Osservatorio 5,
I35122 Padova, Italy. E-mail: Granato@astrpd.pd.astro.it\\
Tel. (049) 8293441, Fax (049) 8759840}
\altaffiltext{2}{International School for Advanced Studies, Strada
Costiera
11, I34014 Trieste, Italy. E-mail: Danese@astrpd.pd.astro.it}
\altaffiltext{3}{Dipartimento di Astronomia, Vicolo Osservatorio 5, I-35122,
Padova, Italy. E-mail: Franceschini@astrpd.pd.astro.it}

\begin{abstract}

We investigate models for the power supply and broad-band spectral
energy distribution (SED) of hyperluminous IR galaxies, recently
discovered at high redshifts, in terms of the emission from an active
nucleus embedded in a torus-like dusty structure.  We find consistent
solutions in terms of a simple torus model extended several hundreds of
parsecs, with $A_V$ in the equatorial plane of a few hundreds and a
typical covering factor of over $50\%$.  Objects as different as the
prototype high-z galaxy \fd, the z=0.93 IR object \fq, \fz\ found in a
high-z cooling flow, and the optically selected BAL "Cloverleaf" quasar,
are all fitted by the same solution, for decreasing values of the polar
angle to the line-of-sight and proper scaling of the luminosities.  We
suggest that such luminous high-z IR objects are heavily buried quasars
surrounded by large amounts of dust with high covering factors and large
optical depths.  Comparison with UVX QSOs suggests that they are observed
during a transient phase.  Forthcoming observations in the far-IR will
soon allow probing this phase and its relationship with the -- possibly
concomitant -- formation of the nuclear black hole and of the host
galaxy.

\end{abstract}


\section{INTRODUCTION}

A new class of peculiar hyperluminous IR galaxies $(L_{I\rm R} / L_\odot
> 10^{12} h^{-2}$, for $q_0=0.5$ and $h = H_o/100$ km s$^{-1}$ Mpc
$^{-1}$) has been recently discovered at moderate to high redshifts.
They appear to have fairly similar luminosities, in particular after
that the most extreme example, \fd \ at $z=2.286$ with an observed
$L_{\rm IR}\sim 3 \times 10^{14} h^{-2} L_\odot$, has been proven to be
significantly amplified by a foreground gravitational lens (e.g.
Broadurst and Lehar, 1995). Other well studied hyperluminous IR objects
are \fq\ ($z=0.93$ and $L_{\rm IR}\sim  1 \times 10^{13} h^{-2}
L_\odot$) and IRAS 09104+4109 ($z=0.442$ and $L_{\rm IR}\sim 6 \times
10^{12} h^{-2} L_\odot$).

While the Cloverleaf is definetely a QSO, the presence of an active
nucleus in the other 3 objects is supported by various observations,
including strong emission lines typical of local type--2 AGNs (e.g.
Soifer et al.\ 1995), the optical-UV polarization (e.g.  Hines et al.
1995), and, concerning in particular the mostly investigated \fd, the
severe constraints on the size of the source (Eisenhard et al. 1995) and
the recently discovered broad emission lines in polarized light (Miller
1995).

Quite an impressive comparison has been recently suggested by
Barvainis et al. (1995) between the overall spectrum of \fd\ and that
of the Cloverleaf radio-quiet BAL quasar at z=2.558, which has an
identical spectrum to \fd\ in the IR/sub-mm, but is two orders of
magnitude brighter in the optical. This is suggestive that \fd\ and
Cloverleaf might be the high redshift and high power counterparts of
narrow- and broad--line AGN, respectively. The only difference
between the two would be in the line-of-sight, intersecting a
torus-like distribution of gas and dust in the former, while falling
close to the almost clean polar cap in the latter.

However, published fits to the IR spectra in terms of emission from dust
illuminated by a luminous AGN fail to reproduce the observed broad dust
emission spectra in the rest-frame 10 to 500 $\mu m$, for all such
objects.

This letter is mainly devoted to show quantitatively that the emission
of a dusty torus--like structure  dominates the observed IR spectra of
hyperluminous IR galaxies harbouring a {\it bona fide} active nucleus.
We use the radiative transfer code developed by Granato and Danese
(1994), which successfully fits the available data on both broad and
narrow lined local AGNs (Granato and Danese, 1994; Granato et al.,
1996).  The possible effects of gravitational lensing in modifying the
shape of the SEDs is properly taken into account. The relationship among
IR and optically selected AGN are also discussed.

\section{THE TORUS MODEL}

A thorough discussion of our adopted dust torus model and the numerical
method developed to compute the emitted SED can be found in Granato and
Danese (1994). Only its basic features are summarized here.

The code solves, through an iterative numerical scheme, for the transfer
equation of the radiation originating from the central optical--UV
source in an axially symmetric dust distribution.

The dust is composed of a mixture of grains, in thermal equilibrium with
the radiation field, and extends out to a maximum radius $r_{max}$.  We
adopted the 6 silicate and graphite grain model described by
Rowan--Robinson (1986) \nocite{Row86} which fits reasonably well the
average galactic absorption law. The inner boundary of the cloud,
$r_{min}$ is set by the sublimation of all species of grains, which is
assumed to happen at temperatures of 1500 K for graphite and 1000 K for
silicate grains.

To keep the number of free model parameters to a minimum, we have
adopted a dust density distribution constant with radial distance from the
central source and function of the polar angle $\Theta$ only:

\begin{equation}
\begin{array}{ll}
\! \! \! \! \rho(\Theta) = C \exp(-\gamma \cos^2\Theta)
\end{array}
\label{equden}
\end{equation}

\noindent for $r_{min} \leq r \leq r_{max}$. This allows to have
directions close to the poles of the distribution less affected by dust
extinction (the classical torus configuration). Such optically thin dust
along the axis scatters a fraction (of order of a few percent) of the
primary spectrum into orthogonal directions, producing a highly
polarized `secondary' spectrum. This distribution is then parametrized
by the three constants $C$, $r_{max}/r_{min}$, and $\gamma$, where the
normalization $C$ is conveniently replaced by the equatorial optical
thickness $\tau_{e}$ of the cloud at 0.3 \micron\ ($A_V\simeq 0.61
\tau_{e}$ and, assuming a standard dust--to--gas ratio, $N_H\simeq
1.2\times 10^{21} \tau_{e}$ cm$^{-2}$).

As for the input blue-bump spectra, we employed a functional description
consisting in a broken power law with $\alpha=-0.5$ for $\log \nu
<15.4$, $\alpha=-1.0$ for $15.4 \le \log \nu <16$ and a sharp cutoff
with $\alpha=-2.2$ for $\log \nu \ge 16$. This provides  a good
approximation for the observed optical--UV portion of the blue bump, and
is consistent with indications coming from observations of high--z
objects, photo-ionization models for the emission lines and computations
of accretion disk spectra for the EUV regime.

\section{MODEL SEDs VERSUS DATA}

Fits to the observed mm-IR-UV spectral energy distributions of the
Cloverleaf, \fd, \fz\ and \fq\ are shown in Figure 1.  {\it The same
model with $\tau_e=250$, $r_m/r_o=500$, and $\gamma=6.2$ is shown here
to account remarkably well for all objects, when observed from different
viewing angles.}

\subsection{The case of IRAS F10214+4724 and the Cloverleaf}

The SEDs of Cloverleaf is reproduced if the torus is observed close to
the pole, while that of \fd\ when it is seen along the equator, provided
that the latter SED is upscaled in luminosity by a factor 2.

The analysis of the Cloverleaf and \fd\ is complicated, however, by the
fact that they are gravitationally lensed objects. For a given lens
geometry the lensing magnification $m$ depends on the source size and
surface brightness distribution. Since the emission in different
spectral regimes arises from regions of different sizes, not only the
normalization but also the shape of the SED could be affected. In
particular, in our tori the extension of the emission usually, but not
always, increases with $\lambda$. The surface brightness distribution
is, in any case, rather complex and heavily dependent on the
line-of-sight and observation wavelength.

We show in Figure 3 contour maps of the model for an edge-on
line-of-sight at four representative wavelengths. For such an equatorial
view, the observed emission of our torus configuration would be rather
extended at any wavelengths. At far-IR wavelengths the system is
optically thin and extends to $r_{max}$. In the near--IR, the hot inner
dust is heavily obscured, while the relatively weak optical--UV
continuum consists of photons initially escaping along directions close
to the polar axis, where the optical thickness is low, and then
scattered along our line-of-sight by dust in the polar region.

On the contrary, when the model is observed from a direction with
moderate line-of-sight extinction (pole-on view), we see directly the
near--IR emission from the hot dust, as well as the primary source.  The
emission then becomes more and more compact at decreasing wavelength.

We have performed a detailed modelling of the lensing geometry for
different lines of sight. The basic results on the chromatic effect of
lensing are summarized here, while for the full treatment we defer to
Granato et al.  (1996).

Though uncertain, the lensing geometries for the Cloverleaf and \fd\ are
clearly much different.  In the former case, the observed quadruple
image is explained by putting the source well inside the tangential
diamond caustic, where the magnification is of order of a few, with a
small gradient. It is weakly dependent on the source size, as long as it
keeps inside the caustics and relatively far from the cusps.  Thus,
despite the strong wavelength dependence of the torus, when observed
from the pole, we find that the predicted magnification is achromatic to
within 20\%.

For two independent reasons, one related to the complex source geometry
(see Figure 2), and the other to the source position with respect to the
diamond caustic, the case of \fd\ is more intricate. The arc observed in K
and optical may be explained by centering the source just outside a cusp of
the diamond. In this region the amplification is large and has a large
gradient. Given the non-circular source simmetry, the magnification depends
also on the angle $\alpha$ between the torus axis in Figure 2 and that of the
diamond caustic in the plane of the sky.

Then, any solution for the lensing geometry and for the chromatism of
\fd\ is model dependent to some extent.  If we adopt the lens solution
of Broadhurst and Lehar (1995) and the torus model of Figure 2, we find an
average magnification factor of $m \simeq 15$ and chromatic distortions
of the SED within factors of 3.  An HST image of \fd\ taken at $0.8$
\micron\ by Eisenhard et al. (1996) provides strong support to the
gravitational lensing hypotesis.  The arc is unresolved in the transverse
direction, giving a direct upper limit to the radius of the UV source of
about 150 $h^{-1}$ pc, consistent with our inferred value of $\sim$ 100
$h^{-1}$ pc for our UV model source (see Figure  2).  On the other hand,
their values for the total magnification $m\simeq 100$ and for the UV
source size of 20 $h^{-1}$ pc inferred indirectly from the flux ratio of
the arc to the counterimage, are made uncertain by differential
extinction, time delays and microlensing along the two light paths. They
also estimate a magnification of a factor about 30 for the IR emission.

Although these two models predict different magnification on very small
scales, nevertheless they agree that the effect of the lensing in
changing the ratio of the optical to far-IR luminosities is within a
factor of 3.  The main effect would be on the relative normalization of
the IR and the optical--UV spectrum in Figure 1, which however could be
easily accounted for with small changes in the equatorial optical
thickness and/or viewing angle.

\subsection{The case of \fq\ and \fz}

The lack of submillimeter observations leaves the SEDs for these two
sources less well defined.  However it is interesting to notice that the
same model, conveniently down--scaled in luminosity, provides a
reasonable fit also to these data when observed from an intermediate
angle $\Theta \simeq 65^o$ (Figure 1).

Obviously, fit parameters for individual sources are rather poorly
constrained (e.g. $r_{max}/r_{min}$ can be changed to 800 and 300,
respectively, and an interplay between the viewing angle and $\tau_e$ is
allowed). But additional observations help in better understanding these
objects. Hines et al.\ (1995) infer an intrinsic spectrum of \fq\
remarkably similar to that of a `typical' QSO, with a luminosity in
close agreement with our model value (cfr. Table 1). In our
configuration, the optical--UV photons are yielded by scattering of the
intrinsic spectrum by dust grains, which is responsible for the observed
polarization (cfr. Hines et al 1995).

High polarization is also observed in \fz\ (Hines \& Wills 1993). The
stellar continuum is important here longward $0.45$ $\mu$m (rest frame)
and diluites the observed polarization (Hines \& Wills 1993; Kleinmann
et al. 1988). After correction for starlight we are left with a
$F\propto \nu^{-0.6}$ continuum with an absolute magnitude close to the
value required by the model. The predicted polarization is also close to
the the observed values, provided that dilution at level of 30--50 $\%$
by stellar light is taken into account.

\section{DISCUSSION}

The observed SEDs of the four hyper-luminous IR objects are consistent
with the presence of dust distributions encompassing $\sim 10^7 M_\odot$
around a luminous AGN, with $A_V \simeq 100-300$ over 50\% or more of
the nuclear sky, $r_{max}/r_{min}$ of several hundreds, and with a
fraction of the remaining sky (the ``cones'') containing more diffuse
matter which scatters some of the primary continuum into our
line-of-sight. Further information on our best-fit solutions is provided
in Table 1.

More compact and thicker tori, as proposed by Pier and Krolik (1993) in
a different context, would produce quite different SEDs, with pronounced
peaks at 8-10 $\mu$m quickly falling at longer $\lambda$ for nearly
face--on objects.  For instance, a torus model with their standard
values of the parameters ($r_{max}/r_{min} \lesssim 10$ and $\tau_{e}
\gtrsim 1000$) would predict a fall--off of the SED (in $\nu L_\nu$) by
more than two orders of magnitude going from the peak down to 100
$\mu$m, whereas the Cloverleaf exhibits a rather flat spectrum up to
$\sim$ 30 $\mu$m with a decrease of less than a factor of ten at 100
$\mu$m (see Figure 1).  In the case of sources viewed at large polar
angles, compact and very thick tori proposed by Pier and Krolik would
produce narrow spectra peaking around 10--15 $\mu$m (whilst the spectra
raported in Figure 1 exhibit maxima at around 30--40 $\mu$m) and with
too steep falls off both at longer and shorter wavelengths.  A detailed
comparison of various torus models may be found in Granato \& Danese
(1994).

An interesting step would now be to understand if this sample of
hyperluminous objects could be interpreted as just the optically thick
population viewed edge-on in a unified picture of quasar activity, where
the commonly UVX selected objects are the pole-on optically thin
counterpart. A clue to this question may come from a comparison of the
Cloverleaf's overall spectrum with those of UVX QSOs and of Seyfert 1
nuclei, as reported in Figure 3.

We see here that the average IR to optical luminosity ratio of Seyferts
1 is higher by a factor of 2-3 than the same for UVX QSOs, a fact that
may be interpreted as an effect of the nuclear luminosity on the
environment, producing a decreasing dust covering factor and/or optical
depth at increasing luminosity. We also see that Cloverleaf is at
variance with respect to this picture, because its IR to optical
luminosity ratio is even higher than that of low-luminosity Seyfert 1
nuclei.

Our model fits to the data summarized in Figure 3 may be characterized,
for example, by the equatorial optical thickness $\tau_e$ of the dust
torus.  A typical SED of high luminosity UVX QSOs is fitted by values in
the range $5<\tau_e<15$, for lower luminosity Seyferts 1 Granato and
Danese (1994) found $10<\tau_e<40$, while for NGC1068 $50\lesssim\tau_e
\lesssim 100$.  For Cloverleaf and the other hyperluminous IR objects we
find here $100<\tau_e<500$.  Our conclusion is that the latter class of
AGNs are strongly at variance with respect to the other usual UVX AGNs,
not only because of the very high inferred values of $\tau_e$, but also
because of their very high nuclear luminosities, which breaks down the
trend for a lower coverage at higher luminsities.

Note, finally, that Cloverleaf exhibits Broad Absorption Lines (BAL) in
the spectrum, a situation that may arise from radiation-pressure
acceleration of material in the torus polar regions for very high,
possibly super-Eddington, accretion rates.  All this is suggestive of a
possible initial transient phase of formation of a super-massive
black-hole, when the gas is still very abundant around the nucleus.

How this phase relates with that of formation of the host galaxy?  The
detection of a large amount of CO in \fd\ may indicate that star
formation is an ongoing process in the host galaxy.  Also the detection
of dust in the sub-millimeter for a few high-z quasars may be in
agreement with the idea of an initial phase in which a `young' nucleus
is immersed in a `young' galaxy with large amounts of gas not yet locked
into stars.

Any conclusion is obviously premature untill a good enough statistics
will be available on this class of sources. The forthcoming ISO mission
is expected to improve it substantially, given its dramatic sensitivity
improvement with respect to IRAS. We estimate that of order of one
thousand UVX AGNs will be discovered in the foreseen 25 sq. deg. sky
area covered by various ISO surveys  at $\lambda \simeq 5$ to $100\ \mu
m$. The dusty QSOs that will be found, whose number is unpredictable now
since we do not know the duration of the dust enshrouded phase, will
cast light on the quasar formation process and its relationship with
that of the host galaxy.

\acknowledgments
We are indebted to G. De Zotti for critical reading of the paper and
fruitful discussions.

\newpage

\begin{table*}
\caption{Model solutions for the bolometric luminosity of the primary
optical-UV continuum $L_{\mbox{bb,46}}$ in units of 10$^{46}$ erg
s$^{-1}$ Hz$^{-1}$, the inner $r_{min}$ and outer radius $r_{max}$ of
the dust torus and total dust mass involved.  For the two lensed
objects, all quantities depend on the assumed magnification factor $m$,
likely to be $\gtrsim 10$ for \fd\ and $\sim 3$ for Cloverleaf.}
\centering
\begin{tabular}{lcccc}
$\phantom{1234567890123}$ & & & \\
\multicolumn{1}{c}{Object} & $L_{\mbox{bb,46}}$ & $r_o[\mbox{pc}]$
&$r_{max}[\mbox{pc}]$ & $M_d[M_\odot]$\\
Cloverleaf\dotfill    & 22 \mmu \hmd & 2.3 \mmum \hmu  & 1170 \mmum \hmu  &
$3.5 \times 10^{7}$ \mmu \hmd\\
\fd     \dotfill    & 40 \mmu \hmd & 3.2  \mmum \hmu & 1580  \mmum \hmu  &
$6.4 \times 10^{7}$  \mmu \hmd \\
\fq     \dotfill    & 4.0  \hmd    & 1.0 \hmu       & 500   \hmu	&
$6.4 \times 10^{6}$  \hmd \\
\fz     \dotfill    & 1.6  \hmd    & 0.6  \hmu     &  316  \hmu	 &
$2.6 \times 10^{6}$  \hmd \\
\end{tabular}
\label{tablr}
\end{table*}

\newpage

\figcaption {A {\it simultaneous fit} of the dust enshrouded QSO model
to the rest-frame SEDs of the four hyper-luminous IR sources.  The
dusty torus has an equatorial optical thickness at 3000 \AA\ of
$\tau_e=250$ ($A_V=150$), the ratio between the outer and inner
radius is 500, and the density depends only on the polar angle (see
Eq. 1), so that at the pole $\tau_p=0.50$.  \fd\ and the
Cloverleaf correspond to an equatorial and polar viewing angle,
respectively.  The SEDs of \fq\ and \fz\, down-scaled in the figure
by a factor 10 and 30 to avoid confusion, are reasonably well
reproduced by the {\it same model}, but viewed from $\Theta \simeq
45^o$ and conveniently scaled in luminosity.  }

\figcaption {Brightness contours at four different wavelengths of the
circumnuclear dust distribution used to reproduce the SEDs of the
Cloverleaf and hyperluminous IRAS galaxies. In this case the ``torus''
is seen edge--on, i.e.\ $\theta=90^o$. The boxes have a width of $2
r_{max}$, and the levels refers to 0.01, 0.03, 0.1, 0.3 and 0.9 of the
peak brightness.}

\figcaption{ Comparison of the average broad-band spectra for various AGN
classes. UVX QSOs (Elvis et al.\ 1994) appear as a dashed, and Seyfert 1 nuclei
(Granato \& Danese 1994) as a dotted line. SED data on Cloverleaf (Barvainis
et al. 1995). The SEDs have been normalized at 18 $\mu$m.}

\newpage

\begin{figure}
\figurenum{1}
\epsscale{1.0}
\plotone{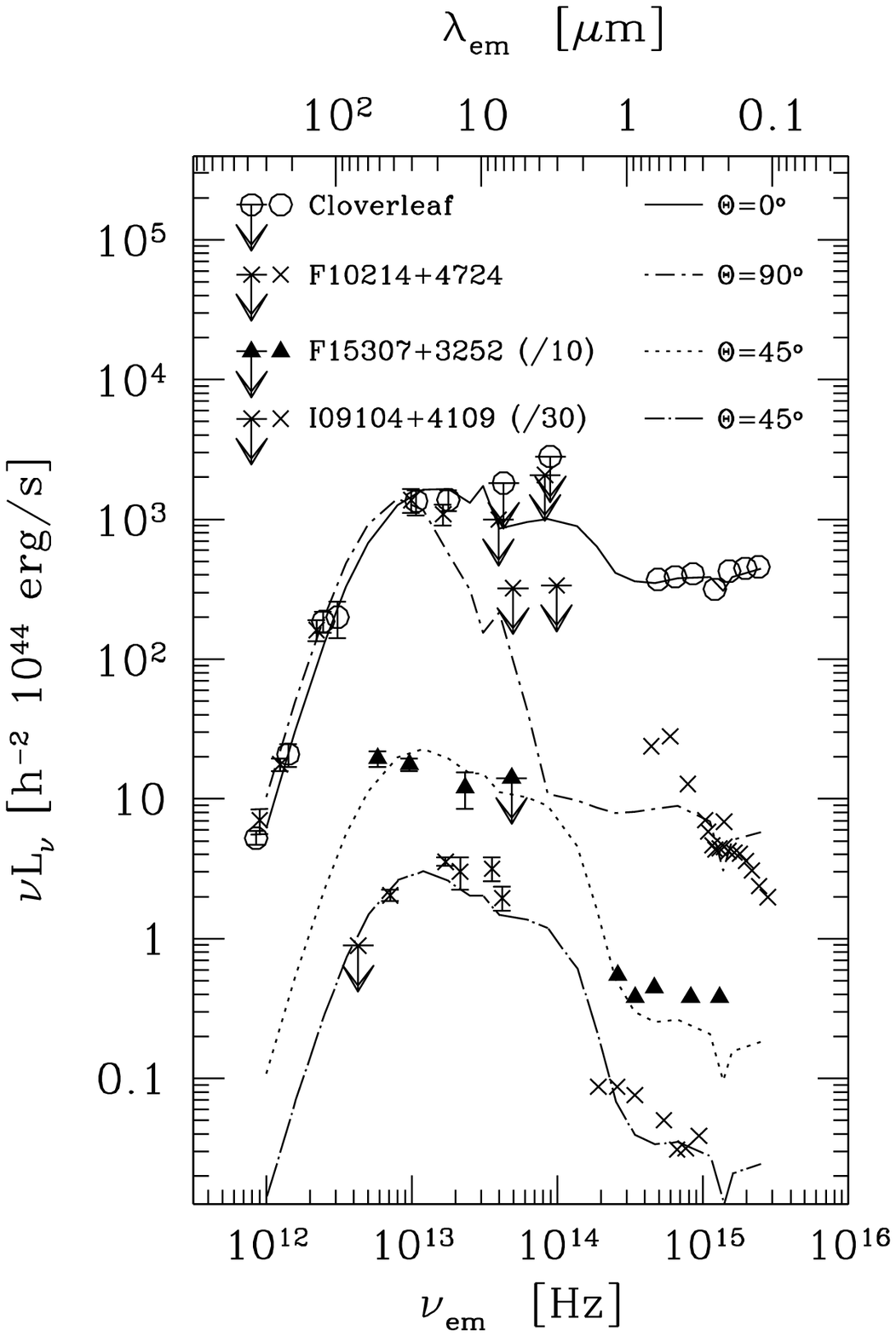}
\end{figure}

\newpage

\begin{figure}
\figurenum{2}
\epsscale{1.0}
\plotone{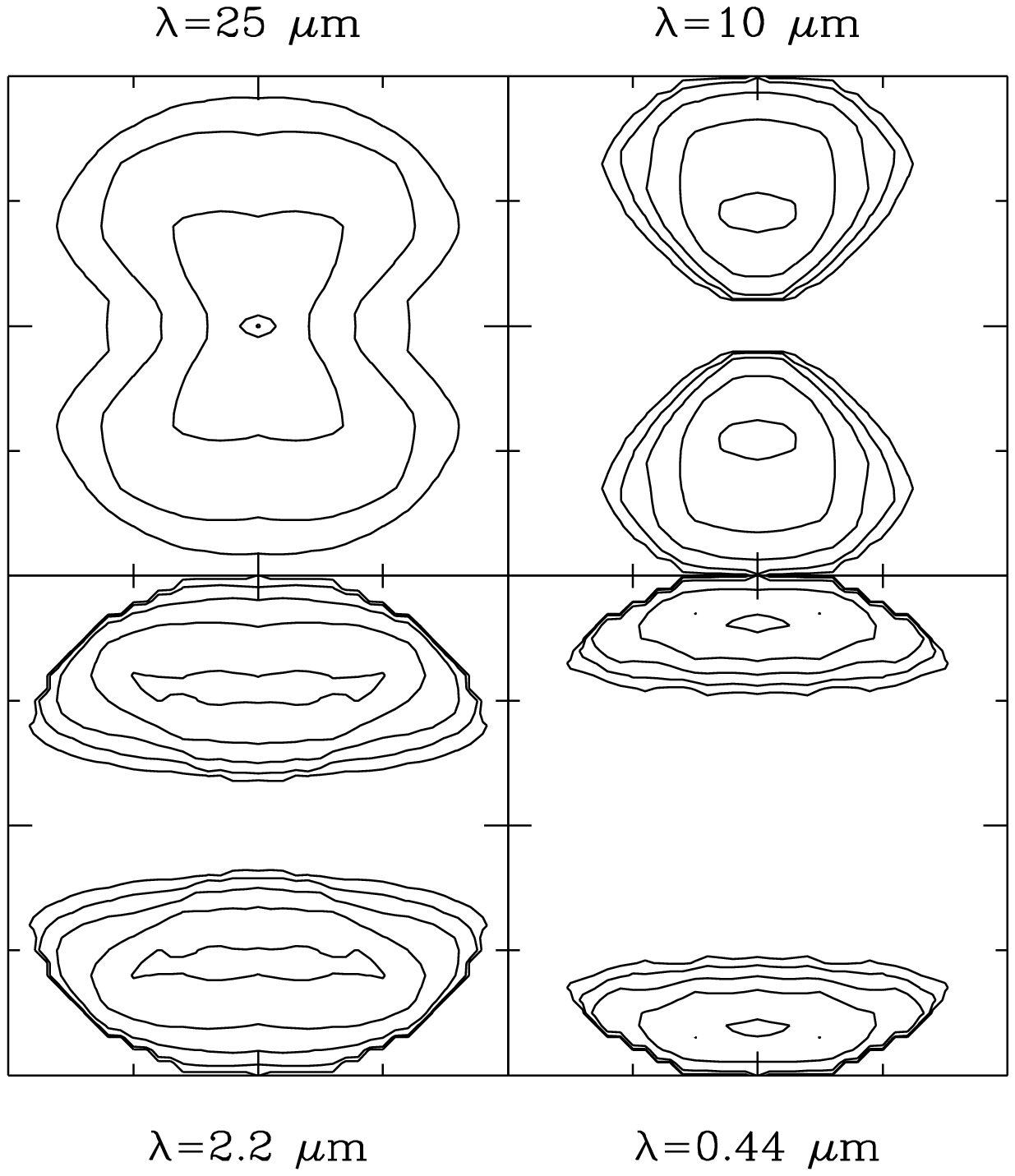}
\end{figure}

\newpage

\begin{figure}
\figurenum{3}
\epsscale{1.0}
\plotone{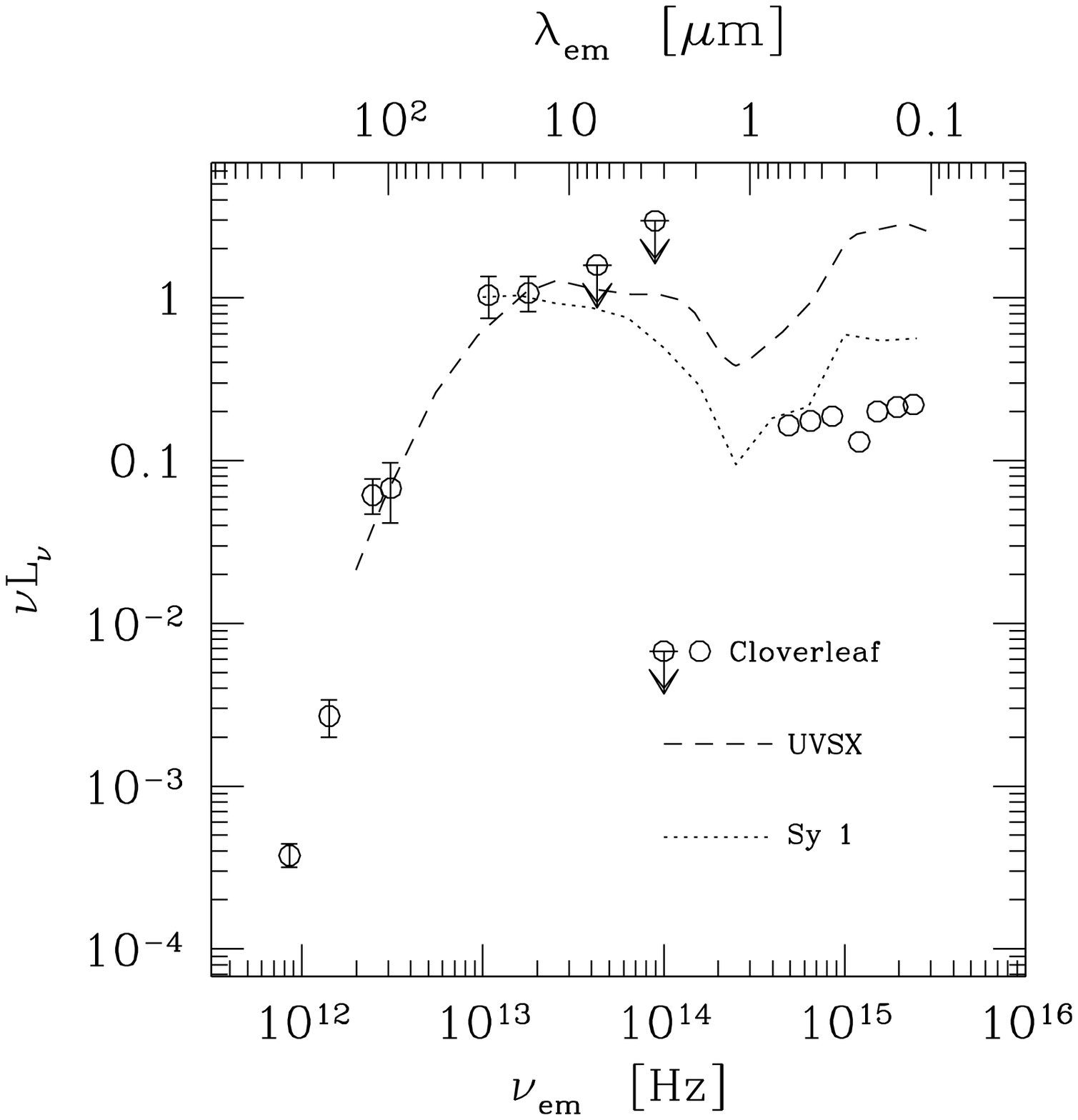}
\end{figure}

\end{document}